\def\year{2018}\relax
\documentclass[letterpaper]{article} 
\usepackage{aaai18}  
\usepackage{times}  
\usepackage{helvet}  
\usepackage{courier}  
\usepackage{url}  
\usepackage{graphicx}  
\usepackage[hide]{chato-notes}

\newcommand\eg{\textit{e.g.,~}}

\newcommand\cf{\textit{cf.~}}
\newcommand\etal{\textit{et al.~}}
\newcommand\etc{\textit{etc.~}}
\newcommand\vs{\textit{vs.~}}
\newcommand{\one}{({\em i}\/) }
\newcommand{\two}{({\em ii}\/) }
\newcommand{\three}{({\em iii}\/) }
\newcommand{\four}{({\em iv}\/) }
\newcommand{\five}{({\em v}\/) }
\newcommand{\six}{({\em vi}\/) }
\newcommand{\seven}{({\em vii}\/) }

\newcommand\gareth[1]{\textbf{\textcolor{blue}{G: #1}}}

\begin{document}
%
 \title{WhatsApp, Doc?\\ A First Look at WhatsApp Public Group Data\thanks{Code and dataset from this paper can be found at https://github.com/gvrkiran/whatsapp-public-groups}}
\author{Kiran Garimella\\EPFL, Switzerland\\kiran.garimella@epfl.ch \And Gareth Tyson\\Queen Mary University, London\\gareth.tyson@qmul.ac.uk}

\maketitle

\begin{abstract}
In this dataset paper we describe our work on the collection and analysis of 
public WhatsApp group data. 
Our primary goal is to explore the feasibility of collecting and using WhatsApp data for social science research. 
We therefore present a generalisable data collection methodology, and a publicly available dataset for use by other researchers. 
To provide context, we perform statistical exploration to allow researchers to understand what public WhatsApp group data can be collected and how this data can be used. 
Given the widespread use of WhatsApp, our techniques to obtain public data and potential applications are important for the community.
\end{abstract}

\section{Introduction}

The Short Message Service (SMS) was initially envisaged as a feature of the GSM standard. It enabled mobile devices to exchange short messages of up to 160 characters. Despite its auxiliary nature, it rapidly became popular; in 2010, 6.1 trillion SMS were sent~\cite{itu_figures10}. However, this is beginning to be surpassed by the emergence of several Internet-based messaging apps, \eg WeChat, Telegram and Viber. 
Although these apps have pockets of dominance, the clear market leader is \textbf{WhatsApp}~\cite{similarweb16}. For example, in India, over 94\% of all Android devices have the app installed with an average of 78\% of current installs using it daily. 

The reasons for its dominance are numerous. Released in 2009, WhatsApp was the forerunner of mobile messaging apps. At this time, many mobile subscribers were charged for sending SMS --- WhatsApp offered a free equivalent, whilst allowing users to maintain many of the convenient aspects of SMS, \eg identification via phone numbers. WhatsApp also introduced powerful new features, such as the ability to include multimedia content and create shared groups. In 2017, WhatsApp reached 1 billion users each day, with 55 billion daily messages being sent~\cite{verve17}.

This suggests that a major portion of online interactions take place via WhatsApp. Indeed, its popularity far exceeds more traditional messaging services likes Skype~\cite{similarweb16}. However, its group functionality and easy integration of multimedia content indicates that usage may differ significantly from these other platforms, particularly SMS. This is confirmed in social studies that have found  that WhatsApp tends to be used in a more conversational and informal manner amongst close social circles~\cite{church2013s}. 
 A particularly novel aspect of WhatsApp messaging is its close integration with public groups. These are openly accessible groups, frequently publicised on well known websites,\footnote{For example, \url{https://joinwhatsappgroup.com/}} and typically themed around particular topics, like politics, football, music, \etc This constitutes a radical shift from the bilateral nature of SMS. 
As such, we argue that these public WhatsApp groups warrant study in their own right. More generally, although past studies have investigated WhatsApp usage via methodologies such as interviews~\cite{church2013s}, we believe it is important to perform both large-scale and data-driven analyses of its usage. 

With this in mind, this \textbf{dataset paper} presents a methodology to collect large-scale data from WhatsApp public groups. To demonstrate this, we have scraped 178 public groups containing around 45K users and 454K messages. Such datasets allow researchers to ask questions like
\one~Are WhatsApp groups a broadcast, multicast or unicast medium?
\two~How interactive are users, and how do these interactions emerge over time?
\three~What geographical span do WhatsApp groups have, and how does geographical placement impact interaction dynamics?
\four~What role does multimedia content play in WhatsApp groups, and how do users form interaction around multimedia content?
\five~What is the potential of WhatsApp data in answering further social science questions, particularly in relation to bias and representability?

We begin by presenting related studies that have either focussed on WhatsApp or messaging services more generally (\S\ref{sec:rw}). 
Due to the difficulty in data collection, most of these studies rely on qualitative methods and interviews/surveys. Our dataset therefore constitutes the first large-scale public WhatsApp data source. We then describe our data collection methodology, which involves scraping a list of public WhatsApp groups, subscribing to them, and then monitoring them such that all communications can be imported into an easy-to-use schema (\S\ref{sec:methodology}). 
With this data, we then proceed to perform a basic characterisation, outlining its key trends (\S\ref{sec:char}). We particularly focus on exploring the potential, as well as the biases we see in the dataset. We conclude that collecting large-scale public messaging data with WhatsApp \emph{is} feasible, and one can obtain a broad geographical coverage (\S\ref{sec:discussion}). However, we also find that diversity amongst groups is high (both in terms of activity levels, geography and topics covered). Hence, a careful selection of seed groups is paramount for meaningful results. In summary:

\begin{itemize}
	\item We show the possibility of collecting publicly available WhatsApp data. 
	\item Using the above approach, we collect an example dataset of 178 groups, containing 45K users and 454K messages.
	\item We characterise the patterns of communication in these groups, focussing on the frequency, types and topics of messages.
	\item We show the applicability of such data in answering new social science research questions.
	\item We release an anonymised version of the data and all the code used to allow others to collect targeted datasets on groups relevant to their research.

\end{itemize}

\section{Related work}
\label{sec:rw}

We see two major themes of related work: \one~studies that have explored social communication patterns on SMS and similar messaging services; and \two~studies that have focussed on WhatsApp itself.

\vspace{6pt}
\noindent\textbf{Studies of Messaging} There have been a large number of studies exploring user behaviour regarding messaging. Due to popularity amongst teenagers, many studies have focused on their usage patterns. This has included work across various countries, including Finland~\cite{kasesniemi200211}, Norway~\cite{ling200210}, the United Kingdom~\cite{grinter2001tngrs,grinter2003wan2tlk,faulkner2004fingers} and the United States~\cite{battestini2010large}. 
Generally, services like SMS have been found to be primarily used within close social groups for activities such as general conversation, planning and coordination~~\cite{grinter2001tngrs}. This is driven by its low cost, ease of use and lightweight nature. Other research has focused on the language used, including the emergence of text-based slang~\cite{grinter2003wan2tlk} and usage of messaging across different age ranges~\cite{kim2007configurations}. A key limitation of these studies has been the focus on qualitative methodologies, \eg interviews, surveys, focus groups. 
One study collected quantitative data via the installation of a logging tool on user devices~\cite{battestini2010large}. By recruiting 70 participants, they analysed 58K sent messages. Although powerful, this approach is largely non-scalable and creates datasets that are challenging for public use due to privacy constraints. 
Other messaging apps, such as WeChat~\cite{huang2015fine}, have been explored at scale although the focus has not been on the content and interactions. Instead, coarser analyses have been performed, \eg size of messages. Studies that have explored more social features have, again, limited themselves to small-scale surveys~\cite{lien2014examining}.
It is worth noting that there have also been several studies exploring messaging patterns within other community mediums, \eg Reddit~\cite{singer2014evolution}, 4chan~\cite{bernstein20114chan,hine2017kek} and IRC~\cite{rintel2001first}. We consider such platforms orthogonal to WhatsApp, and therefore do not focus on them here.

\vspace{6pt}
\noindent\textbf{Studies of WhatsApp} There have been a small number of studies that have inspected the usage of WhatsApp specifically. Due the differences between WhatsApp and SMS, these deserve discussion in their own right. 
These studies tend to centre on WhatsApp usage within given settings. For example, there have been studies inspecting how students and teachers interact via WhatsApp~\cite{bouhnik2014whatsapp}, as well as the impact WhatsApp usage may have on school performance~\cite{yeboah2014impact}. Similar studies have been performed within medical settings to understand how WhatsApp facilitates communication amongst surgeons~\cite{wani2013efficacy,johnston2015smartphones}. The common limitation of these studies is their reliance on small populations and qualitative methodologies (\eg interviews). Although important, this provides little insight into more general purpose usage across ``typical'' users. 
Church \etal also performed a direct comparison of SMS \vs WhatsApp, finding that interviewees used WhatsApp more often, confirming its growing importance~\cite{church2013s}. 

In contrast to the above studies, which rely on surveys and interviews, \cite{rosenfeldwhatsapp} took a quantitative approach by harvesting WhatsApp data directly from 92 volunteers. Due to the private nature of the messages, the authors focused on metadata rather than message content, \eg length of text. Montag \etal took a similar approach, asking 2418 users to download an app that records usage~\cite{montag2015smartphone}. 
Both works are highly complimentary to our own; the main difference is that we focus on \emph{public} rather than private WhatsApp communications, allowing us to yield datasets with orders of magnitude more users. This is because the intrusive nature of the data collection in these other studies makes it difficult to scale-up beyond small numbers of users.

\section{Data collection}
\label{sec:methodology}
This section delineates the data collection methodology, as well as its limitations and ethical considerations. Both the tools and datasets are publicly available.

\subsection{Data Collection Methodology}

We begin by detailing our data collection methodology. We intend this to be generalisable across any set of WhatsApp groups or, indeed, other online messaging services that support public groups. For this, we only required a single low-capacity compute server, alongside a working mobile device with WhatsApp installed. A single working phone number is required, such that the WhatsApp SMS confirmation can be received to register the device. Once these tools are in place, the data collection contains two steps.

\vspace{6pt}
\noindent\textbf{Step 1} 
First, it is necessary to acquire a set of public groups for data collection. We are not prescriptive in how these are obtained. For example, some researchers may wish to manually curate a list or target just a small number of highly specific groups. This is supported by a number of existing websites that index public groups (\eg \url{joinwhatsappgroup.com/}). We, however, took a more large-scale approach. We used the Google search engine, and other focussed websites, to compile a list of public groups. This was attained by searching for links that contain the suffix of \url{chat.whasapp.com}.\footnote{An example of a public WhatsApp group: \url{https://chat.whatsapp.com/BZp0Ye2eoRp2TWnQe7ixvO}} This gave us a list of 2,500 groups.

Next, we randomly sampled 200 groups from this list and joined them using an automated script.
The script uses a browser automation tool, Selenium and the 
\url{web.whatsapp.com} web interface
to automate the joining process. 
Note that the web interface needs a single time sign in (via scanning a QR code) with the same account as the Android device we will use to subsequently collect the data. 
At the conclusion of Step 1, we had a dedicated WhatsApp account subscribed to the full set of groups with little human intervention in the process. Hence, this can easily scale to much larger sets of groups.

\vspace{6pt}
\noindent\textbf{Step 2}
Once we joined the groups, we started to receive updates on the phone. As WhatsApp implements end-to-end encryption\footnote{https://www.whatsapp.com/security/} it is naturally difficult to passively collect data on the device (\eg via Wireshark). 
Fortunately, WhatsApp stores all messages received within a simple sqlite database on the local device. This made it trivial to extract the data being collected periodically from the device (once the storage began to fill). 
To make this feasible, however, it was necessary to use the encryption keys to decrypt the stored version of the messages.\footnote{Messages are both transmitted and stored in an encrypted form} 
We therefore used the technique of of Gudipaty \etal~\cite{gudipaty2015whatsapp} to extract the storage key and decrypt all messages.\footnote{The encryption key can also be obtained in a much simpler manner with a rooted Android phone, e.g. see \url{http://jameelnabbo.com/breaking-whatsapp-encryption-exploit/}.}
Overall, we collected data for 178 groups,\footnote{22 out of the 200 groups were either removed or had no activity.} containing 45,794 users, and 454,000 messages over a 6 month period (May-Oct 2017).

The anonymised dataset and code are publicly available.\footnote{\url{https://github.com/gvrkiran/whatsapp-public-groups}}

\subsection{Ethical Considerations}

Clearly the above methodology has the capacity to collect large bodies of data containing messages sent by individuals from around the world. There are therefore certain privacy considerations that must be taken into account. Most notably, individual phone numbers should \emph{not} be collected and/or released. To anonymise users, we allocate each phone number a unique identifier after extracting the appropriate country code. We also advice researchers to delete the WhatsApp device database after data has been extracted from the device (because the WhatsApp database will continue to store the phone number). To further guarantee privacy, we also do not release message content in our public dataset (just metadata). Researchers are welcome to contact us for bespoke access to the full dataset.  

Researchers should also be careful regarding which types of groups they choose to scrape. Although all groups are public and therefore users are aware that their messages will be seen by unknown parties, it is worth noting that there are a wide diversity of group types. These include those of an adult nature, which some researchers may wish to avoid, \cf \cite{tyson2015} for further discussion. Moreover, researchers will have no control over the content sent via the groups; hence, there is a risk of receiving unsavoury or even illegal multimedia content. Our advice is therefore to disable the automatic downloading feature on the device running WhatsApp (this is also helpful for improving scalability).

Finally, we emphasise that the privacy policy for WhatsApp groups states that a user shares their messages and profile information (including phone number) with other members of the group (both for public and private groups).\footnote{\url{https://www.whatsapp.com/legal/}}
Group members can also save and email upto 10,000 messages to anyone.\footnote{\url{https://faq.whatsapp.com/en/android/23756533/}} 
Our paper provides automated tools for this process. 


\section{Characterising WhatsApp groups}
\label{sec:char}

To provide context for the applicability of WhatsApp group data, we next characterise its basic properties. We particularly focus on identifying the issues and biases that may occur within such data. Although we utilise our collected dataset to underpin this, other researchers can apply a similar methodology to acquire data in their target domains.

\subsection{How much data can be collected?}

\begin{figure}[t]
\centering
\includegraphics[width=0.5\textwidth]{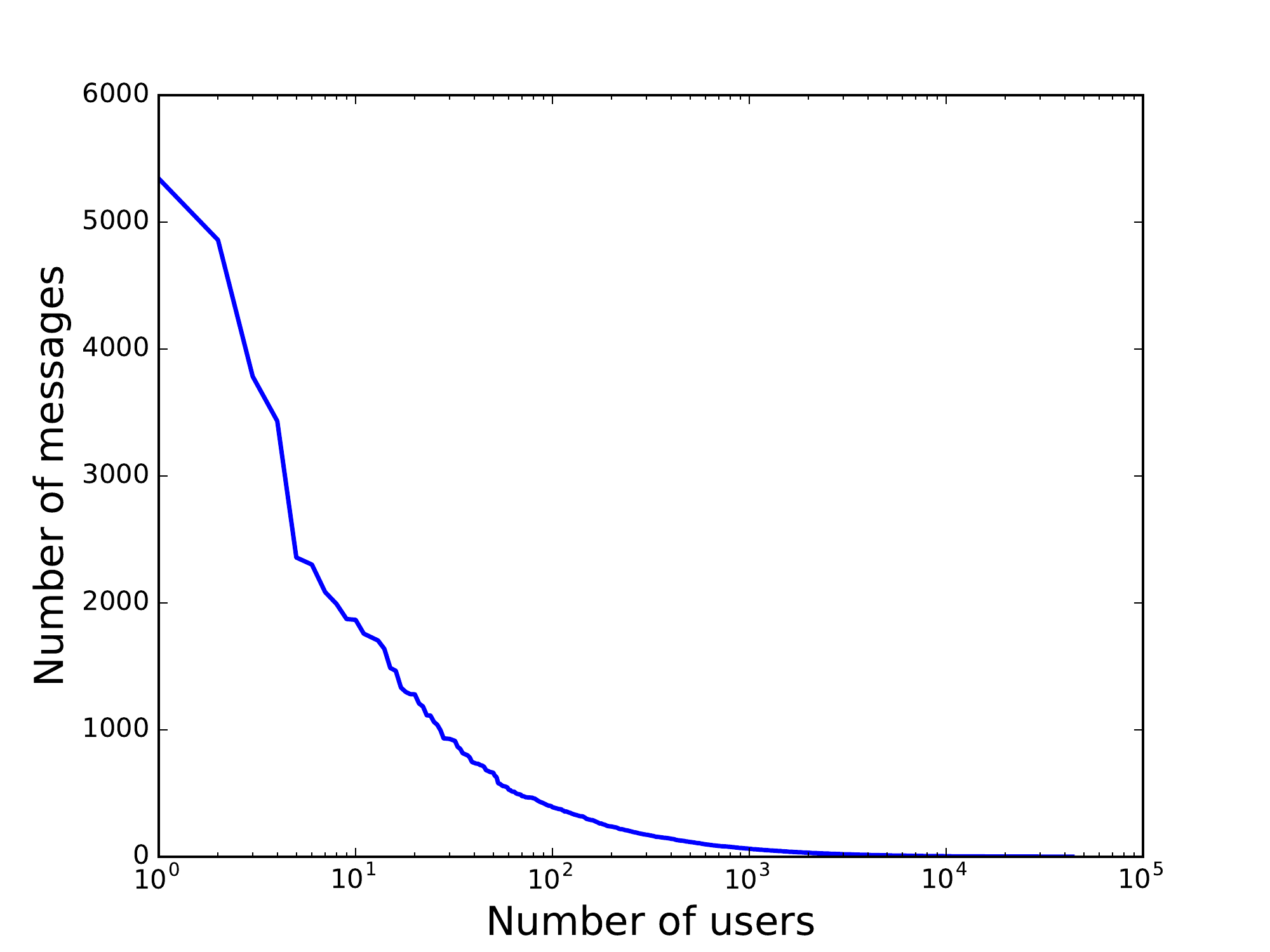}
	\caption{\label{fig:users_activity}Activity of users in our dataset. 75\% of the users have less than 5 messages.}
\end{figure}

Over the 6 month period, we collected data from 178 groups. Each group had an average of 143.3 participants (median 127), with the largest group observed containing 314 participants.\footnote{Note, at the time of writing there is a default maximum of 256 group members per group, which can be increased manually.} 
In total, 454K messages were collected, spanning 45K users. Figure~\ref{fig:users_activity} presents the number of messages sent per-user. Unsurprisingly, the distribution is highly skewed with the top 1\% of users generating 37\% of all messages. 
Around 10K users (25\%) have more than 5 messages. The remaining 75\% of the users are mostly consumers of information.

\begin{figure}[t]
\centering
\includegraphics[width=0.5\textwidth]{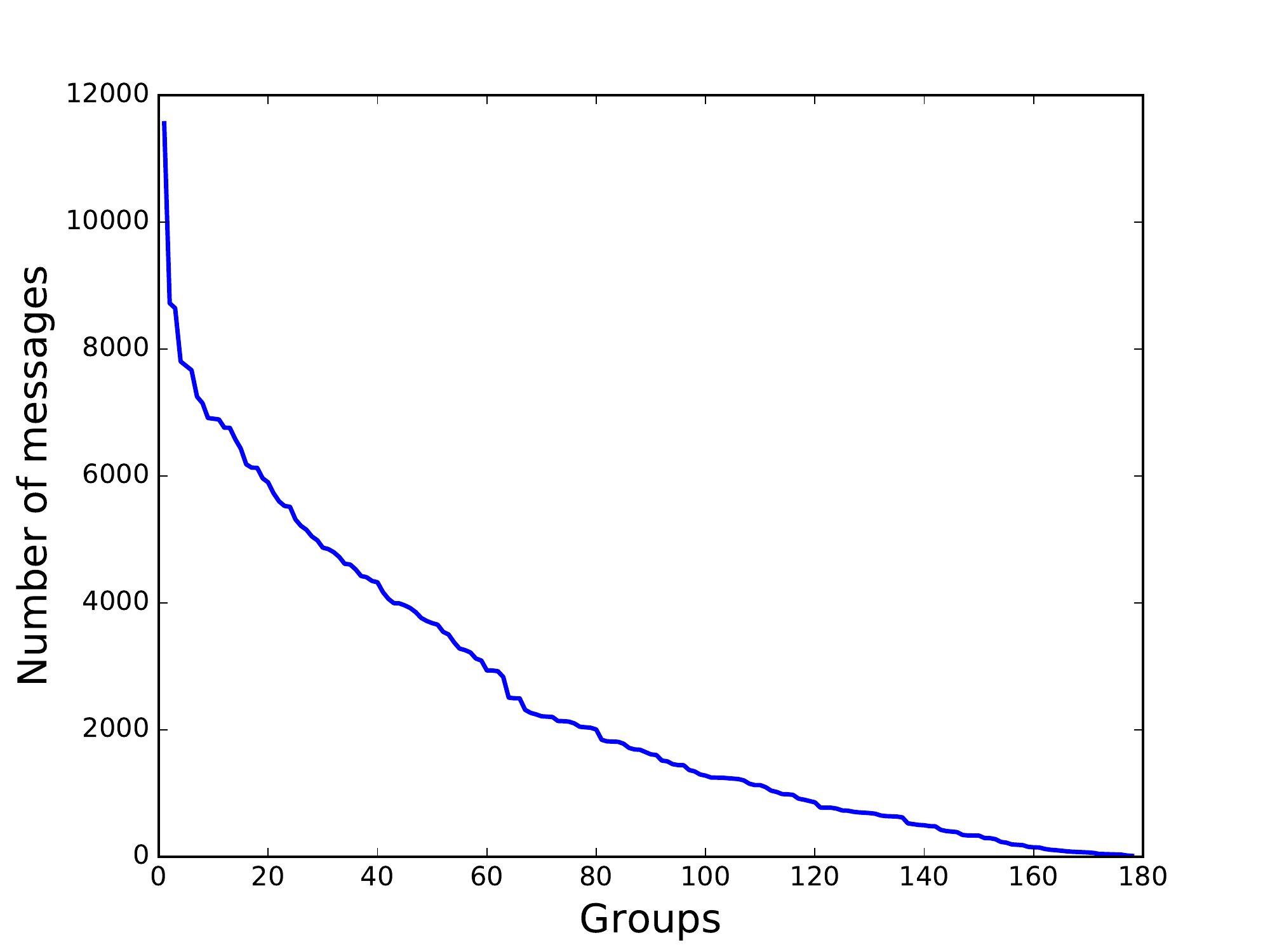}
	\caption{\label{fig:groups_activity}Number of messages per group. 
	Over 30\% of the groups have less than a 1000 messages in 6 months.}
\end{figure}

Figure~\ref{fig:groups_activity} shows how these messages are distributed across groups. We find that over 30\%  of the groups have under 1000 messages during the 6 month measurement period. 
Despite this, there are a small number of highly active groups --- the most active generated 11K messages overall. This indicates there is a high degree of scope for optimisation with researchers being able to get significant volumes of messages from just a few groups. Data from the top 10 groups would yield in excess of 80K messages (18\% of our overall set). 
As such, it is clear that WhatsApp can be effectively used for garnering significant social datasets.

\subsection{Where are users located?}

The above has shown that large quantities of social data \emph{can} be collected from WhatsApp groups. We next ask what geographical biases may be contained within such data.
Each user is associated with a phone number. By examining the country code, it is possible to geolocate users based on their registered country. This has the benefit of not changing whilst users are visiting other countries (unlike datasets based on GPS or IP geolocation). 

\begin{figure}[t]
\centering
	\includegraphics[width=0.5\textwidth,clip=true,trim=10 200 10 100]{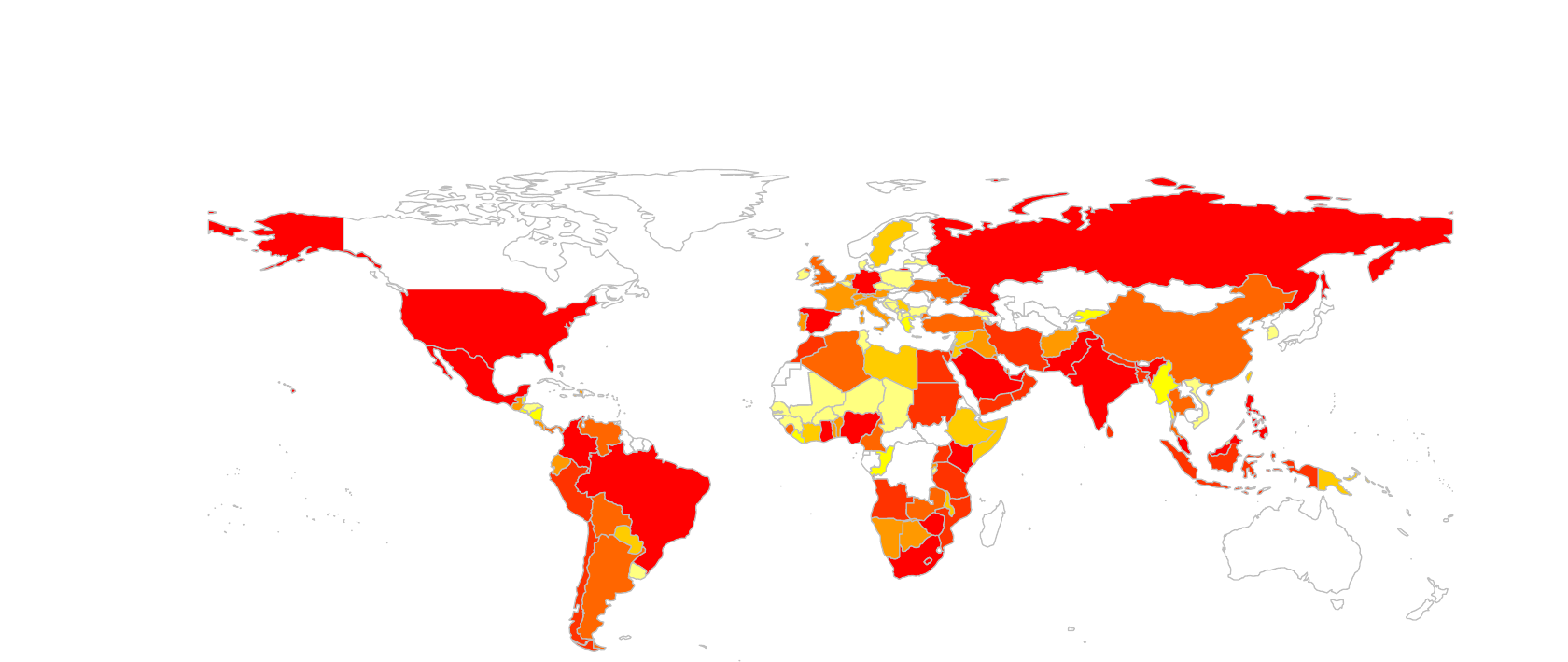}
	\caption{\label{fig:users_location} Location of users in our dataset. Brighter shades of red indicate higher number of users.}
\end{figure}

Figure~\ref{fig:users_location} presents a heatmap of user locations.
The top countries include India (25K), Pakistan (3.6K), Russia (3K), Brazil (2K) and Colombia (1K). 
This immediately confirms a significant geographical bias, although not towards the United States as one would typically expect. 
This may therefore be considered as a positive point by many social science researchers. 
For example, we see many users in developing regions, \eg in Africa, Nigeria has 959 users, whilst in South America, Colombia has 1,073 users. 
Hence, we posit that these datasets may offer effective cultural vantage into developing regions as well as developed ones. 

\begin{figure}[t]
\centering
\includegraphics[width=0.5\textwidth]{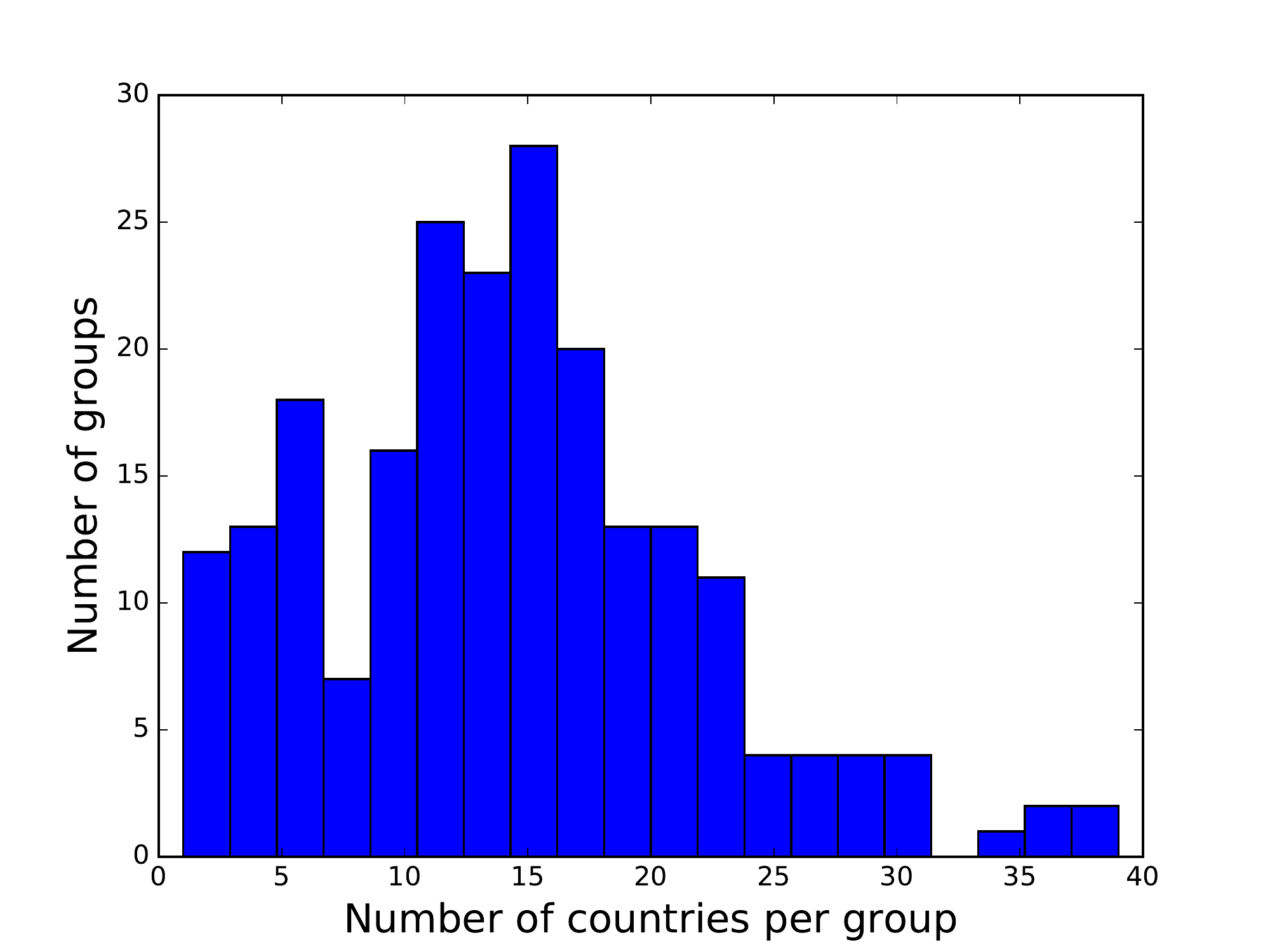}
	\caption{\label{fig:groups_countries_hist}Histogram showing number of countries users in a group belong to. A majority of the groups have users from more than 10 countries.}
\end{figure}

This diversity is also mirrored in the make-up of individual groups. Remarkably, we do not find \emph{any} groups that are limited to a single country. 
Instead, all groups contain members from multiple countries. Figure~\ref{fig:groups_countries_hist} presents a histogram of the number of countries contained within each group. It can be seen that significant international communities are present within the groups. 
85\% of groups have members from over 10 different countries. Again, this indicates that the data offers a vantage into globalised communities that easily cross national boundaries. 
We looked at the 5 groups that have users from more than 30 countries, to find that they varied in type, including sex, English learning, YouTube videos, \etc

Another property of geography is language. We automatically inferred the language of a message using Lui \etal~\cite{lui2011cross}. Note that our analysis on language depends on the performance of their model.
Across the 178 groups, we observe 59 languages which have at least 200 messages sent.
Table~\ref{tab:languages} presents a breakdown of the most popular languages. Unsurprisingly, English is most prominent with in excess of 137K messages. This is followed by Hindi, and other Indian languages such as Gujarati, Tamil and Marathi. Although a powerful feature in itself, this does significantly complicate analysis. 
Unfortunately, many groups contain messages of multiple languages, making deeper social analysis even more challenging. 
This is not just occasional messages as we find that 33\% of groups have less than 50\% of messages in a single language. 

\begin{table}[t]
\centering
\begin{tabular}{ l | l  }
\hline
\# Messages & Language \\
\hline\hline
137527 & English\\ 
78333 & Hindi \\
13063 & Spanish \\
7525 & Gujarati\\
5341 & Tamil\\
5123 & Chinese\\
4193 & Marathi\\
2942 & German\\
2930 & Polish\\
2349 & Italian\\
\hline
\end{tabular}
\caption{Top 10 most popular languages as measured by number of messages sent.}
\label{tab:languages}
\end{table}

\subsection{What is sent?}

We now progress to explore the content of what is sent within the groups. 
We remind the reader that this is heavily impacted by the choice of groups being scraped. 
As previously stated, we collected 454K messages overall. From these, 9.1\% were images, 3.6\% were videos, and 0.7\% audio; the rest were text. The average image size is 101KB, whilst the average video is a non-negligible 4.6MB. The average length of the text messages 582 characters (median 136 characters). 

As well as content, we observe a large number of URLs being shared --- a remarkable 39\% of messages contain web links. This offers a powerful tool for researchers wishing to explore social web content popularity. Table~\ref{tab:domains} presents the most popular domains shared via WhatsApp, as well as their Alexa Ranking. 
Although we observe many of the international hypergiants (\eg Google, YouTube) we also observe a wide range of fringe websites. There is little correlation between the popularity of the domain in our WhatsApp data and its popularity on Alexa. 
Of course, this is partly driven by the geographical distribution of the user base; for example, \url{lootdealsindia.in} has a global rank of 917,011 but an Indian ranking of 83,911. 
Despite this, it is clear that WhatsApp groups may offer an effective vantage into lesser known web content and how it is accessed by fringe communities. 

\begin{table}[t]
\centering
\begin{tabular}{ l | l | l }
\hline
\# Messages & Domain & Alexa Rank\\
\hline\hline
59883 & youtube.com & 2\\
37270 & whatsapp.net & 614,880 \\
12239 & amazon.in & 90 \\
7141 & google.com & 1\\
5395 & whatsapp.com & 69\\
3979 & blogspot.com & 63\\
1989 & wowapp.com & 78,514 \\
1218 & flipkart.com & 161 \\
1144 & lootdealsindia.in & 917,011 \\
1032 & marugujraat.com & 6,217,479 \\
952 & kamalking.in & 799,769\\
630 & dealvidhi.com & 2,895,020\\
455 & facebook.com & 3 \\
453 & mydealone.com & 7,882,171 \\
431 & msparmar.in & 5,008,742 \\
405 & newsdogshare.com & 163,914 \\
402 & newsdesire.com & N/A\\
346 & sex.xxx & N/A \\
324 & ojasinfo.com & 2,949,092 \\
323 & jobdashboard.in & 324,811 \\
\hline
\end{tabular}
	\caption{Most popular domains within URLs shared via WhatsApp groups. whatsapp.net urls mostly contain multimedia. google.com is mostly for sharing playstore apps (play.google.com).}
\label{tab:domains}
\end{table}

\begin{figure}[t]
\centering
\includegraphics[width=0.5\textwidth]{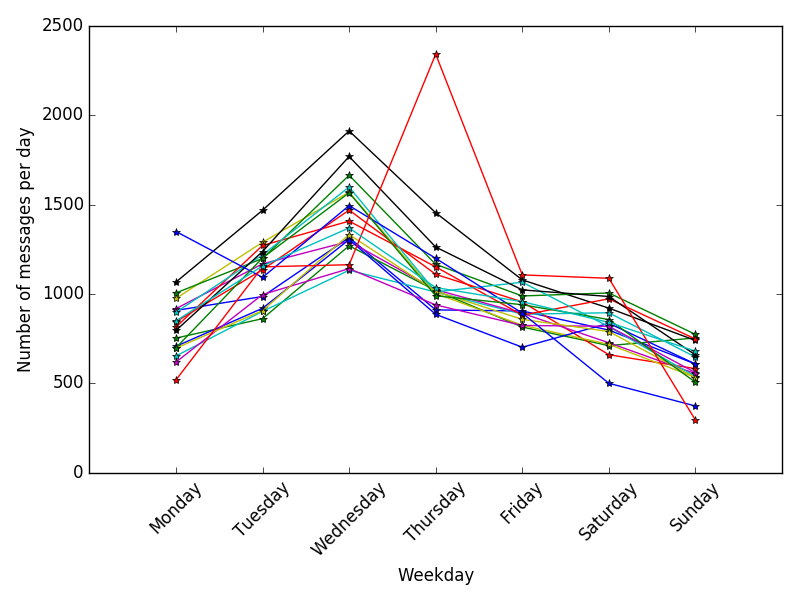}
	\caption{\label{fig:groups_weekday_activity}Number of messages sent per day for the top 20 groups with highest activity.}
\end{figure}

We can also inspect the temporal trends of when these messages are sent. Figure~\ref{fig:groups_weekday_activity} depict the total number of messages sent on each day of the week for the top 20 groups in terms of activity. 
Two noteworthy things can be observed. First, the greatest activity occurs on weekdays, rather than weekends. 
Second, the peak day for most groups is Wednesday. Why this might be is unclear, however, it is evident that this holds across many groups. 79\% of all the 178 groups peak on a Wednesday.
This trend is in line with other social networks like Facebook and Twitter, where previous studies have revealed increased activity during weekdays with peaks on Wednesday.\footnote{\url{http://bitly.tumblr.com/post/22663850994/time-is-on-your-side}}
It is also worth briefly noting that very few (under 2\%) of these messages are replies.\footnote{Users can directly send replies to other messages} This is a feature that is rarely used, therefore making it difficult for researchers to formally understand who is talking to whom within groups.

\subsection{What topics are captured?}

Finally, we inspect the topics captured within the groups. There is no formal taxonomy of topics within WhatsApp and, thus, it is necessary for researchers to manually inspect and classify the groups under study. We manually annotated the 178 groups we collected into a set of categories. 
From our WhatsApp dataset, we find several types of groups with significant followings:
\one  \emph{generic groups} -- `funzone', `funny', `love \vs life', \etc (70 groups);
\two \emph{adult groups} -- `XXX', `nude', \etc (19 groups);
\three \emph{political aligned groups} -- mostly Indian political parties (15 groups);
\four \emph{movies/media} --- `box office movies', fan groups, anime, etc (17 groups);
\five \emph{spam} --- deals, tricks (14 groups);
\six \emph{sports} --- football (`football room'), cricket (`world cricket fans'), \etc (12 groups);
\seven other -- job posts, education discussion, tech, activism, \etc (23 groups);
 
\begin{figure}[t]
\centering
\includegraphics[width=0.5\textwidth]{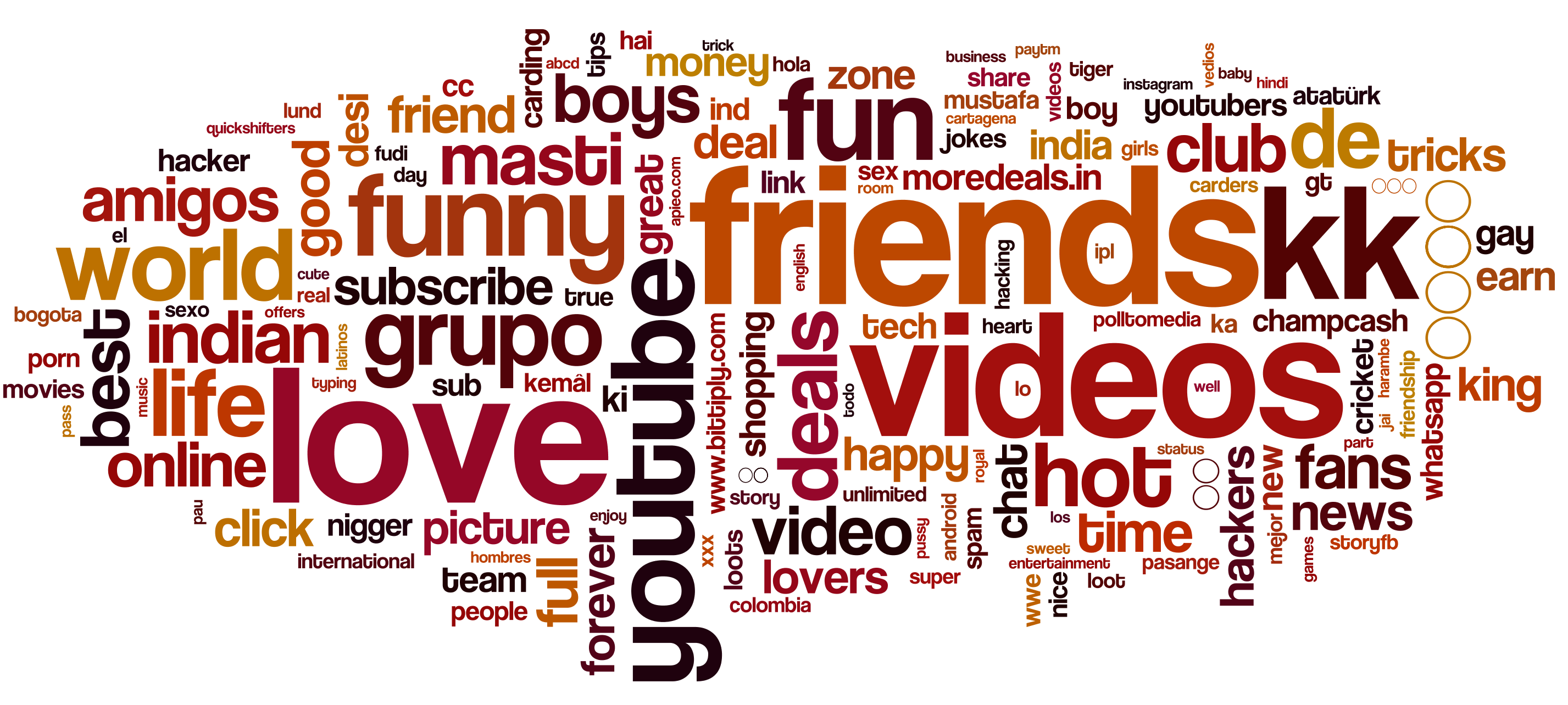}
	\caption{\label{fig:all_groups_wordcloud}Word cloud generated from group titles. All 2500 groups identified in Step 1 of the methodology were used.}
\end{figure}

Hence, researchers wishing to focus on any of these topics could certainly do so via WhatsApp data.
The largest group is ``DISFRUTA AL MAXIMO'' (enjoy to the fullest) which contains 11K messages, primarily based in Colombia, followed by ``No life without cricket'' (8.7K messages, India), and ``Football room'' (7.7K messages, Nigeria). Again, we emphasise that these statistics are biased by our choice of groups, however, their diversity confirms that it would be possible for many different topics to be explored via these groups.
Briefly, to provide finer-grained vantage of the topics discussed, we can inspect the words used within the group titles. Figure~\ref{fig:all_groups_wordcloud} presents a word cloud generated using the group titles.
In-line with the above topics, we observe regular discussions related to concepts such as nudity, videos and cash, as well as geographical indicators such as India.

\section{Conclusion \& Discussion}
\label{sec:discussion}

The paper has provided tools to collect WhatsApp data for the first time.
The dataset we collected is a random sample of 178 public groups, however, the principle behind this paper is to show that large scale data collection from WhatsApp groups is feasible.
Such datasets, if collected with a predefined goal in mind, have immense consequences and open up new areas of research.

As well as presenting our methodology, we have also performed a basic characterisation of our dataset to highlight its key features. This has revealed potential bias in factors such as geographical user distribution. However, rather than being a limitation, we believe such bias could be exploited. For example, one important finding is the ability to collect data both globally and across borders. 
Although this naturally covers highly connected regions such as Europe and North America, we also observe a significant number of users in developing regions. 
Thus, we argue that WhatsApp may be particularly useful for offering vantage into such regions (which are often overlooked in mainstream research).
For example, in India alone, it is estimated that by 2020, 400 million new users who have never been a part of the digital data realm, will join the Internet.
The popularity of WhatsApp means that it could act as a powerful research tool for understanding this growing use. 
With this in mind, we conclude by listing a few ambitious questions that we believe WhatsApp group data may be able to help answer:



\begin{enumerate}
\item  Can we find the emergence of new social institutions from WhatsApp group data? Given this new ecosystem of connectivity that empowers users, new institutions such as markets (micro work, virtual trading), money (\eg WeChat money, AliPay, PayTM), and social organisations (trade unions) may emerge. How would such trends be reflected in WhatsApp activity? 

\item  Can we understand the role of these new institutions in shaping the economic, social and wellbeing of the people who constitute these institutions? For instance, understanding the effects of new markets on patterns of migration and assimilation between villages and cities. WhatsApp data could potentially expose these patterns as users come and go between groups, and  as new groups emerge to reflect these institutions.

\item Can we use this data to explore and understand how information such as ``fake news'' spreads through communities. This is particular relevant as fake news is a significant issue on WhatsApp, especially in countries with low levels of digital literacy.\footnote{\url{http://bit.ly/2DuStFn}}
More generally, how does multimedia content propagate through (and spread between) such groups?

\item  Can we make use of the insights taken from WhatsApp groups to create algorithms to help deliver better services to users, which can improve their way of life? For example, \one~\emph{Livelihood}: micro-matching jobs and talents, \two~\emph{Wellbeing}: using WhatsApp-shared image analysis for automated medical diagnoses, \three~\emph{Education}: Delivering the right content to the right people --- educating farmers with crop season information, \etc Each of these topics could benefit from their implementation over WhatsApp, \eg using groups to share relevant employment information in communities.


\end{enumerate}

The above topics go well beyond the scope of this initial work. However, as a popular medium for communication in many parts of the world, we argue that WhatsApp should be given equal attention to that of other social media services, \eg Twitter. We hope that this work, and its associated tools, can act as a platform for other research to build atop of.

\bibliographystyle{aaai}
\bibliography{biblio}

\end{document}